# Q-learning-based Hierarchical Cooperative Local Search for Steelmaking-continuous Casting Scheduling Problem

Yang Lv[1], Rong Hu [(1, 2, *)], Bin Qian [(1, 2, *)], Jian-Bo Yang[3]

*Abstract*—The steelmaking–continuous casting scheduling problem (SCCSP) is a critical and complex challenge in modern steel production, requiring the coordinated assignment and sequencing of steel charges across multiple production stages. Efficient scheduling not only enhances productivity but also significantly reduces energy consumption. However, both traditional heuristics (e.g., two-stage local search) and recent meta-heuristics often struggle to adapt to the dynamic characteristics of practical SCCSP instances. To address these limitations, this paper introduces a novel Q-learning-based hierarchical cooperative local search framework, termed HierC_Q, aimed at minimizing the weighted sum of the maximum completion time and the average waiting time in SCCSP. The core contributions of HierC_Q are twofold. First, considering the intrinsic coupling properties of the SCCSP, a dedicated reward function is proposed based on a novel coupling measure (CM), guiding the exploration process towards promising regions of the solution space. Second, a hierarchical architecture is devised, comprising two distinct tiers: the "learn to improve" (L2I) tier and the "disturb to renovate" (D2R) tier. The L2I tier performs deep exploitation within promising regions using two independent Q-learning-based local search frameworks (QLSFs) tailored for subproblems, along with a synergy QLSF designed for the main problem. To enhance the effectiveness of local search, a validity evaluation approach and a speed-up evaluation method are also introduced, grounded in a detailed study of the problem's structure. Meanwhile, the D2R tier incorporates a perturbation-and-construction-based solution renewal strategy to mitigate the risk of premature convergence. The superiority and effectiveness of HierC_Q are demonstrated through extensive comparisons with eleven local search frameworks and nine state-of-the-art algorithms.

*Index Terms*—Steelmaking and continuous casting, reinforcement learning, Q-learning, meta-heuristic algorithm, production scheduling, manufacturing systems.

## I. Introduction

MANUFACTURING systems are crucial in numerous practical applications [1, 2]. Particularly, the steel industry faces fierce competition due to increasing demands from infrastructure projects like high-rise buildings [3], bridges [4], and public transportation systems [5]. Steel production involves sequential stages including raw material preparation, steelmaking, continuous casting, and rolling, each presenting unique challenges [6-8]. The steelmaking and continuous casting (SCC) stage, identified as the primary bottleneck, has thus attracted significant research attention.

The key objective in SCC is the scheduling of molten steel charges across production stages [9]. However, factors such as extreme temperatures, heavy materials, and strict no-wait constraints between stages complicate decision-making. Efficient schedules must meet stringent quality and timing requirements. Thus, the SCC scheduling problem (SCCSP) represents an important and complex research area.

Recent studies primarily propose various SCCSP models and algorithms, yet overlook intrinsic problem structures, resulting in limited computational efficiency. This paper revisits a representative SCCSP from [10] to uncover its inherent structure, informing algorithm development. SCCSP decomposes into two subproblems: charge scheduling and cast scheduling, exhibiting characteristics akin to the strongly NP-hard no-wait hybrid flow shop scheduling problem (NHFSP). Detailed complexity analyses are available in supplementary material. Traditional mathematical programming (e.g., mixed-integer and dynamic programming) is typically impractical for SCCSP due to computational demands [11], [12]. Consequently, meta-heuristic algorithms, valued for flexibility and adaptive search, dominate current approaches [13].

Meta-heuristics iteratively explore and exploit solution spaces by combining heuristic concepts [14], typically modeled as permutation problems [15]. However, SCCSP differs significantly, consisting of coupled subproblems (charges and casts), greatly enlarging its solution space. For example, scheduling 60 charges grouped into 5 casts yields an enormous solution space (5! × 60!), far exceeding simpler scheduling problems. Thus, standard meta-heuristics struggle to efficiently identify optimal solutions.

Insights from SCCSP's inherent properties highlight that closely coupled charge-cast sequences yield high-quality solutions, forming a small but optimal solution subspace. Standard meta-heuristics' broad exploration limits their ability to focus efficiently on this promising region, emphasizing the need for targeted search strategies.

Exploitation strategies also critically impact meta-heuristic performance [16–18]. Local search frameworks (LSFs), incorporating neighborhood operators and iterative greedy search (e.g., VNS [19], ILS [20], ALNS [21], hyper-heuristics [22]), are widely adopted. Recently, reinforcement learning (RL), particularly Q-learning, has emerged as effective for adaptive LSFs due to intelligent search guidance without extensive meta-optimization [23–25]. Thus, we

Manuscript received XXXX XX, XXXX. This research was supported by the National Natural Science Foundation of China (U24A20273 and 62173169), and the Construction Project of Higher Educational Key Laboratory for Industrial Intelligence and Systems of Yunnan Province (KKPH202403003). (Corresponding author: Rong Hu and Bin Qian)

Yang Lv is with the Shanghai Research Institute for Intelligent Autonomous Systems, Tongji University Shanghai 200092, China.

Rong Hu and Bin Qian are with the School of Information Engineering and Automation, Kunming University of Science and Technology, Kunming 650500, China (e-mail: ronghu@vip.163.com, bin.qian@vip.163.com).

Jian-Bo Yang is with the Alliance Manchester Business School, The University of Manchester, Manchester M13 9SS, U.K.

select Q-learning-based local search frameworks (QLSFs) for our study. However, conventional LSFs are unsuitable for SCCSP due to its coupled subproblems. Simple decomposition strategies are insufficient without co-evolutionary mechanisms. Additionally, existing LSFs often neglect solution diversity, becoming trapped in local optima.

To address these challenges, we propose the Q-learning-based Hierarchical Cooperative Local Search (HierC_Q) framework, structured as follows. First, HierC_Q leverages SCCSP's coupling properties through a novel Coupling Measure (CM), guiding exploration toward tightly coupled, high-quality solutions. This represents the first integration of RL with coupling metrics for targeted search. Second, we construct a hierarchical exploitation architecture with two tiers: Learn to Improve (L2I) and Disturb to Renovate (D2R). L2I uses problem-specific neighborhood operators (PDNOs) and cooperative QLSFs—Charge_QLSF, Cast_QLSF, and their synergy SQLSF—to enhance search efficiency. Charge_QLSF employs a validity evaluation method, while Cast_QLSF uses a speed-up strategy to improve computational efficiency. The D2R tier incorporates perturbation mechanisms to maintain search diversity. Third, the proposed CM approach and hierarchical framework extend beyond SCCSP, offering insights applicable to other decomposable scheduling problems.

The paper proceeds as follows: Section II defines the problem model and complexity analysis. Section III introduces the proposed coupling measure. Section IV details the HierC_Q framework and initialization heuristics. Section V presents comparative experimental results and analyses. Section VI concludes with future research directions.

## II. Problem Descriptions

This paper investigates the steelmaking and continuous casting scheduling problem (SCCSP) under the assumption that all charges follow an identical route: steelmaking → refining → continuous casting. In this permutation-based SCCSP model, each charge $k$ ($k=1, 2, \cdots, N$) s sequentially undergoes processing from the initial steelmaking stage ($i=1$) through subsequent refining stages to the final refining stage ($i=S-1$). Each stage has $m_i$ parallel machines, and transportation times $t_i$ between stages are considered. In the continuous casting stage ($i=S$), $N$ charges are grouped into $Z$ casts based on a predefined production scheme, each cast requiring a single operation preceded by a sequence-independent setup time. Non-overlapping constraints ensure that each machine processes at most one charge or cast simultaneously. Minimizing waiting times is critical to reducing thermal losses and avoiding costly reheating. Additionally, charges within each cast must adhere to a predetermined processing sequence.

To highlight the computational complexity of the SCCSP, we formally prove its strong NP-hardness. Given the strict no-wait constraints, transportation times, and sequence-independent setup times inherent in molten iron processing, SCCSP is modeled as a no-wait hybrid flow shop scheduling problem ($F_S(P_{m(i)})|no\text{-}wait, TT, SIST| C_{max}$). We establish NP-hardness via a polynomial-time reduction from the known strongly NP-hard single-machine scheduling problem ($1||C_{max}$) to a simplified no-wait hybrid flow shop problem ($P_m|no\text{-}wait|C_{max}$), subsequently extending this complexity result to the SCCSP. The reduction entails constructing a job-processing scenario with specifically chosen processing times and employing convergent series properties to maintain optimality conditions (refer to Theorem 1 in **Part 2** of the Supplementary Material). Thus, this study explicitly establishes SCCSP's NP-hardness, addressing a significant gap in the existing literature.

The mathematical formulation includes Eqs. (1)-(10). Specifically, Eqs. (1)-(2) define the forward decoding process for calculating completion times. Constraints (3)-(5) eliminate cast breaks in the continuous casting stage. Reverse decoding to minimize total waiting time is modeled in Eqs. (6)-(7). Eq. (8) computes the maximum completion time, while Eq. (9) evaluates the average waiting time. These are combined into a weighted objective function (Eq. (10)), transforming the original multi-objective problem into a single-objective optimization problem. Reflecting industrial priorities, weights $\Psi_1=10$ and $\Psi_2=1$ are used, adjustable based on specific requirements. Further details on the SCC process and decoding strategy are provided in the Supplementary Material.

TABLE I
NOTATIONS APPLIED IN OPTIMIZATION MODEL

| Symbol | Description |
|---|---|
| $M_s$ | Set of the machine contained in stages $i$, $m_i$ is the total number of machines in stages $i$, $M_s=\{m_1, m_2, …m_i\}$; |
| $\pi_k$ | The $k$-th charge in the charge subsequemce $u$; |
| $\pi_j^{cast}$ | The $j$-th cast in the cast subsequemce $v$, $\pi_j^{cast} = \{l_{j-1}+1, l_{j-1}+2, …, l_j\}$, where $\pi_{j-1}+b$ is the $b$-th charge included in the $\pi_j^{cast}$; |
| $\pi$ | The schedule or permutation, $\pi=(u, v)$. |
| $\alpha_j$ | The total number of the charge contained in cast $j$, $\alpha_j=\|\pi_j^{cast}\|$; |
| $\pi_k^{i(a)}$ | Charge $k$ on the machine $a$ in stage $i$; |
| $t_i$ | The transportation time for a charge to be transported from stage $i$-1 to stage $i$; |
| $ST_j$ | Independent setting time for the cast $j$; |
| $T_i(a)$ | The total number of charges or casts on machine $a$ in stage $i$; |
| $p_{\pi_i^{s(a)}}$ | The processing time of $\pi_i$ at machine $a$ in stage $s$; |
| $L_{\pi_k^{s(a)}}$ | The starting time of $\pi_k$ at machine $a$ in stage $s$; |
| $C_{\pi_k^{s(a)}}$ | The completion time of $\pi_k$ at machine $a$ in stage $s$; |
| $C_{max}(\pi)$ | The maximum completion time of schedule $\pi$; |
| $f_{wait}(\pi)$ | The average waiting time of schedule $\pi$; |
| $\Psi_1$ | The weight coefficient of $C_{max}(\pi)$; |
| $\Psi_2$ | The weight coefficient of $f_{wait}(\pi)$; |

$$C_{\pi_k^{1(a)}} = \sum_{b=1}^{k} p_{\pi_b^{1(a)}}, a=1,2,…m_1, k=1,2,…,T_1(a) \quad (1)$$

$$C_{\pi_k^{i(a)}} = \max(C_{\pi_k^{i-1}} + t_i, C_{\pi_{k-1}^{i(a)}}) + p_{\pi_k^{i(a)}} \quad (2)$$

$$\Delta = C_{\pi_{l_{j-1}+b+1}^{S(a)}} - p_{\pi_{l_{j-1}+b+1}^{S(a)}} - C_{\pi_{l_{j-1}+b}^{S(a)}}, \ b=1,2,K,\alpha_j \quad (3)$$

$$\begin{cases} C_{\pi_{l_{j-1}+b+1}^{S}} = p_{\pi_{l_{j-1}+b+1}^{S}} + C_{\pi_{l_{j-1}+b}^{S}}, & \Delta \leq 0; \\ C_{\pi_{l_{j-1}+b}^{S}} = C_{\pi_{l_{j-1}+b+1}^{S}} - p_{\pi_{l_{j-1}+b+1}^{S}}, & \Delta > 0; \end{cases} \quad (4)$$

$$C_{\pi_j^{cast(a)}} = \max(C_{\pi_{l_{j-1}+1}^{S(a)}} - p_{\pi_{l_{j-1}+1}^{S(a)}}, C_{\pi_{j-1}^{cast(a)}} + ST_{\pi_j^{cast(a)}}) + p_{\pi_j^{cast(a)}} \quad (5)$$

$$C'_{\pi_k^i} = C_{\pi_k^i} + \min\{L_{\pi_{k+1}^i} - C_{\pi_k^i}, L_{\pi_k^{i+1}} - C_{\pi_k^i}\}, i=1,…,S-1 \quad (6)$$

$$L'_{\pi_k^i} = C'_{\pi_k^i} - p_{\pi_k^i} \quad (7)$$

$$C_{max}(x) = \max_{a=1,2,K,m_s} \sum_{j=1}^{T_S(a)} C_{\pi_j^{cast(a)}} \quad (8)$$

$$f_{wait}(x) = \sum_{i=2}^{S} \sum_{k=1}^{N} (L_{\pi_k^{i(a)}} - L'_{\pi_k^{i-1(a)}} - p_{\pi_k^{i-1(a)}}) \quad (9)$$

$$\min f(x) = \psi_1 C_{max}(x) + \psi_2 f_{wait}(x) \quad (10)$$

## III. Coupling Analysis And Coupling Measurement

This section analyzes the coupling characteristics inherent in the steelmaking and continuous casting scheduling problem (SCCSP), introducing a Coupling Measure (CM) to quantify these relationships.

### A. The coupling characteristics of SCCSP

The coupling in SCCSP arises primarily from two constraints: (1) Continuous casting can only commence after all charges in the corresponding cast complete the final refining stage. (2) To minimize costly reheating due to thermal loss, the start of each cast must closely follow the completion of its respective charges. Practically, this implies alignment between the priorities of charges and their corresponding casts.

Coupling thus refers to how well the cast and charge subsequences align, significantly influencing the solution space. The solution space size, characterized by $Z! \times N!$, highlights the complexity introduced by coupling. Optimal solutions typically reside in tightly coupled subspaces, which, although small, represent regions of high alignment. Exploring such focused subspaces is critical, a topic largely unexplored in existing literature.

### B. Coupling measurement (CM)

To address this gap, we propose the Coupling Measure (CM), a mathematical approach to quantitatively assess the coupling between cast and charge subsequences. First, we define a fuzzy relation matrix as follows:

*Definition 1 (Fuzzy relation matrix)*: Let $U=\{x_1, x_2,\ldots, x_Z\}$ represent charges in sequence $u$, and $V=\{y_1, y_2,\ldots, y_Z\}$ denote positions. The membership value $\mu_R(x_i, y_j)=r_{ij}\in[0,1]$ represents the likelihood of charge $x_i$ occupying position $j$, forming the fuzzy relation matrix $\boldsymbol{R}^{charge}=[r_{ij}]_{m\times n}$.

For any feasible solution $\pi_{current}=(u_{current}, v_{current})$, it is nontrivial to directly evaluate, or even estimate, the degree of coupling between the two subsequences—especially in large-scale instances. To bridge this gap, a virtual charge subsequence $u_{virtual}$ is introduced, constructed from the cast subsequence $v_{current}=(\pi_1^{cast}, \pi_2^{cast},\ldots, \pi_Z^{cast})$. This virtual sequence is formed by arranging all the charges within each cast consecutively according to their processing priorities, as illustrated in Eq. (11). Here, $l_0+1$ denotes the first charge in $\pi_1^{cast}$, and $l_1$ is the last charge in the same cast.

Let $r(l_{e-1}+k, j)$ denote the membership value of the $k$-th charge in cast $e$ appearing at position $j$ in the solution space ($j=1,\ldots, N$). This value reflects the probability of that charge occupying position $j$. The position of this charge in the virtual sequence is denoted as $Pos(l_{e-1}+k)$. To model the membership function, we adopt the Gaussian function [27], which provides a smooth probability distribution. The Gaussian-based membership function is defined in Eq. (13), where $\sigma$ denotes the variance parameter that controls the sharpness of positional relevance. Using this formulation, the fuzzy relation matrix in Eq. (14) is derived to represent the discrepancy between $u_{current}$ and $u_{virtual}$. Finally, the coupling measure for the current solution $I_{current}=(u_{current}, v_{current})$ is computed using Eq. (15). A detailed example illustrating the Coupling Measure can be found in **Part 3** of the Supplementary Material.

$$u_{virtual}=\{l_0+1,\ldots,l_1,\ldots,l_{e-1}+k,\ldots,l_e,\ldots,l_Z\} \quad (11)$$

$$\sum_{e=1}^{Z-1}\sum_{k=1}^{\alpha_e}(l_{e-1}+k)=N \quad (12)$$

$$r(l_{e-1}+k, j)=\exp\{-\frac{[Pos(l_{e-1}+k)-j]^2}{2\sigma^2}\} \quad (13)$$

$$\boldsymbol{R}_{charge}=\begin{bmatrix} r_{(l_0+1,1)} & r_{(l_0+1,2)} & \cdots & r_{(l_0+1,N)} \\ r_{(l_0+2,1)} & r_{(l_0+2,2)} & \cdots & r_{(l_0+2,N)} \\ \cdots & \cdots & \cdots & \cdots \\ r_{(l_Z,1)} & r_{(l_Z,2)} & \cdots & r_{(l_Z,N)} \end{bmatrix} \quad (14)$$

$$CM=\frac{1}{N}\sum_{i=1}^{N}r(u_{current}(i),i) \quad (15)$$

## IV. HIERC_Q FOR THE SCCSP

This section presents a detailed implementation of the proposed HierC_Q framework, including its core components and the underlying design motivations.

### A. The problem-dependent neighborhood operators

The quality and diversity of neighborhood operators are critical for improving the exploitation capacity of meta-heuristic algorithms [10], [17], [18]. In this work, we design eight problem-dependent neighborhood operators (PDNOs) for the charge subproblem and adopt three common operators for the cast subproblem, enabling effective local search for both.

#### 1) The PDNOs for the charge subproblem

Empirical results show that neighborhood operations with varying positional distances yield distinct effects. Small-scale operators typically enable quick, localized improvements, while large-scale operators can induce significant structural changes over a broader solution landscape. To harness these complementary effects, eight PDNOs are developed for the charge subproblem, based on three basic types—Swap, Insert, and Exchange—and tailored to the specific structure of SCCSP.

Two new categories are proposed: Variable-distance operators: Extended from Swap and Insert, allowing flexible adjustments at different positional ranges; Variable-scale operators: Based on Exchange, supporting solution changes at various scales. These designs enable an adaptive balance between exploration and exploitation during the search. Detailed descriptions of each PDNO are provided in the following subsections.

The *Swap* operator randomly relocates two positions $i$ and $j$ in the charge permutation and charges between these two positions shift along. Where $F_{SNS}^1(u)$, $F_{MNS}^1(u)$ and $F_{LNS}^1(u)$ are a neighbor of charge subsequence $u$ constructed by randomly swapping two charges $\pi_i$ and $\pi_j$ with large size, medium size, and small size.

$F_{SNS}^1(u)=SNS(u,\pi_i,\pi_j),\{i,j\in 1,2,\ldots,N, i\neq j \text{ and } 0<|i-j|\leq N/6\}$;

$F_{MNS}^1(u)=MNS(u,\pi_i,\pi_j),\{i,j\in 1,2,\ldots,N, i\neq j \text{ and } N/6<|i-j|\leq N/2\}$;

$F_{LNS}^1(u)=LNS(u,\pi_i,\pi_j),\{i,j\in 1,2,\ldots,N, i\neq j \text{ and } N/2<|i-j|<N\}$.

The *Insert* operator randomly select a charge to insert into another random position in the charge permutation. Similar to *Swap* operator, three kinds of *Insert* operations with large size, medium size, and small size are designed. Where $F_{SSI}^1(u)$, $F_{MSI}^1(u)$ and $F_{LSI}^1(u)$ are a neighbor of charge subsequence $u$ constructed by randomly inserting the charge $\pi_j$ before the charge $\pi_i$.

$F_{SSI}^1(u)=SSI(u,\pi_i,\pi_j),\{i,j\in 1,2,\ldots,N, i\neq j \text{ and } 0<|i-j|\leq N/6\}$;

$F_{MSI}^1(u)=MSI(u,\pi_i,\pi_j),\{i,j\in 1,2,\ldots,N, i\neq j \text{ and } N/6<|i-j|\leq N/2\}$;

$F_{LSI}^1(u)=LSI(u,\pi_i,\pi_j),\{i,j\in 1,2,\ldots,N, i\neq j \text{ and } N/2<|i-j|<N\}$.

In addition, two types variable scale *Exchange* operators *NE*(1) and *NE*(3) are designed to constructed its neighborhood. The *NE*(1) operator randomly select a charge $\pi_i$, and then two charge $\pi_{i-1}$ and $\pi_{i+1}$ exchange along. The *NE*(3) operator randomly select a charge $\pi_i$, then three pairs of charges $(\pi_{i-1}, \pi_{i+1})$, $(\pi_{i-2}, \pi_{i+2})$ and $(\pi_{i-3}, \pi_{i+3})$ are exchanged separately. Where $F^1_{NE(1)}(u)$ and $F^1_{NE(3)}(u)$ are a neighbor of charge subsequence $u$ constructed by executing the variable scale *Exchange* operators.

$$F^1_{NE(1)}(u) = Exchange(u, \pi_{i-1}, \pi_{i+1}), i \in (1,2,...,N);$$

$$F^1_{NE(3)}(u) = \{Exchange(u, \pi_{i-1}, \pi_{i+1}), Exchange(u, \pi_{i-2}, \pi_{i+2}),$$
$$Exchange(u, \pi_{i-3}, \pi_{i+3})\}, i \in \{1,2,...,N\}.$$

*2) The neighborhood operators for the cast subproblem*

Different from the charge subproblem, three commonly used neighborhood operators (i.e., *Swap*, *Insert*, *Exchange*) are employed for the cast subproblem. Since we have found that there is less difference in the performance of neighborhood operators with different distances for the cast subproblem. The neighborhood operators for the cast subproblem are described as follows.

$$F_{Swap}(v) = Swap(v, \pi_i^{cast}, \pi_j^{cast}), \{i, j \in (1,2,...,Z), i \neq j\}$$

$$F_{Insert}(v) = Insert(v, \pi_i^{cast}, \pi_j^{cast}), \{i, j \in \{1,2,...,Z\}, i \neq j\}$$

$$F_{Exchange}(v) = Exchange(v, \pi_{i-1}^{cast}, \pi_{i+1}^{cast}), \{i \in \{1,2,...,Z\}\}$$

### B. Q-learning-based local search framework (QLSF)

In most local search frameworks (LSFs), it is challenging to identify the most effective neighborhood operator under dynamic search conditions, and a fixed operator sequence rarely performs well across diverse instances. To overcome this, we propose a Q-learning-based local search framework (QLSF) that adaptively selects the most suitable operator according to the current search context.

The learning performance of QLSF relies on four main components: state definition, action set, reward strategy, and action selection mechanism.

(1) **Definition of states**: In traditional Q-learning formulations, states generally represent the characteristics of the external environment. Prior work on Q-learning-based metaheuristics typically defines states based on the degree of convergence and solution-related quality metrics [2], [28], [29]. However, the SCCSP poses unique challenges due to its multi-constraint nature and vast solution space. Directly encoding solution information into the state representation risks causing the so-called curse of dimensionality.

To mitigate this, a neighborhood-based state aggregation technique is employed. This method partitions the state space into several non-overlapping subsets, each corresponding to a specific neighborhood structure, thereby reducing the dimensional complexity while retaining meaningful distinctions between search contexts. Accordingly, for the charge subproblem, the state space is defined by the eight neighborhood structures introduced in Subsection IV-A-1, as follows:

The set of state space of charge subproblem is composed by the eight neighborhood structures proposed in Subsection IV-A-1) as follows:

$$S_{charge} = \begin{cases} s_1^{charge} = F_{SNS} & s_2^{charge} = F_{MNS} & s_3^{charge} = F_{BNS} \\ s_4^{charge} = F_{SNI} & s_5^{charge} = F_{MNI} & s_6^{charge} = F_{BNI} \\ s_7^{charge} = F_{NE(1)} & s_8^{charge} = F_{NE(3)} \end{cases} \quad (16)$$

For the cast subproblem, the set of state space $\Sigma_{cast}$ is composed by the three neighborhood structures proposed in Subsection IV-A-2):

$$S_{cast} = \{s_1^{cast} = F_{Swap} \quad s_2^{cast} = F_{Insert} \quad s_3^{cast} = F_{Exchange}\} \quad (17)$$

Considering the vastness of search space and the complexity of constraints, a state combination technique is applied to construct the joint state space of SCCSP.

$$S_{joint} = [s_{i,j}^{joint}]_{3 \times 8} = \begin{cases} (s_1^{charge}, s_1^{cast}) & (s_2^{charge}, s_1^{cast}) & ... & (s_8^{charge}, s_1^{cast}) \\ (s_1^{charge}, s_2^{cast}) & (s_2^{charge}, s_2^{cast}) & ... & (s_8^{charge}, s_2^{cast}) \\ (s_1^{charge}, s_3^{cast}) & (s_2^{charge}, s_3^{cast}) & ... & (s_8^{charge}, s_3^{cast}) \end{cases} \quad (18)$$

(2) **Definition of actions**: an action is defined as a state transition from one neighborhood structure to another by executing the improvement operators of the late one.

The action space of charge subproblem $A_{charge}$ and cast subproblem $A_{cast}$ is composed by the neighborhood operators proposed in Subsection IV-A:

For the charge subproblem, the action set $A_{charge}$ is composed by the eight neighborhood operators

$$A_{charge} = \begin{cases} a_1^{charge} = SNS(u) & a_2^{charge} = MNS(u) & a_3^{charge} = BNS(u) \\ a_4^{charge} = SNI(u) & a_5^{charge} = MNI(u) & a_6^{charge} = BNI(u) \\ a_7^{charge} = NE(1)(u) & a_8^{charge} = NE(3)(u) \end{cases} \quad (19)$$

$$A_{cast} = \{a_1^{cast} = Swap(v) \quad a_2^{cast} = Insert(v) \quad a_3^{cast} = Exchange(v)\} \quad (20)$$

Similar to the construction method of the joint state space, an action combination technique is applied to build the joint action space of SCCSP in Eq. (22).

$$A_{joint} = [a_{i,j}^{joint}]_{3 \times 8} = \begin{cases} (a_1^{charge}, a_1^{cast}) & (a_2^{charge}, a_1^{cast}) & ... & (a_8^{charge}, a_1^{cast}) \\ (a_1^{charge}, a_2^{cast}) & (a_2^{charge}, a_2^{cast}) & ... & (a_8^{charge}, a_2^{cast}) \\ (a_1^{charge}, a_3^{cast}) & (a_2^{charge}, a_3^{cast}) & ... & (a_8^{charge}, a_3^{cast}) \end{cases} \quad (21)$$

To facilitate visualization of the state transfer process, a fully connected digraph network $G=(V(G), E(G))$ is shown in Fig.1, where $V(G)$ is a non-empty vertex set, each vertex represents a state (i.e., a neighborhood structure). $E(G)$ is an edge set in which each arc connecting two states represent the action. For the example in Fig. 4, the initial state is $s_1^{charge}=F_{SNS}$, we can perform improvement operator *MNS* on the charge subsequence $u$ of the candidate solution to transfer to $s_2^{charge}=F_{MNS}$.

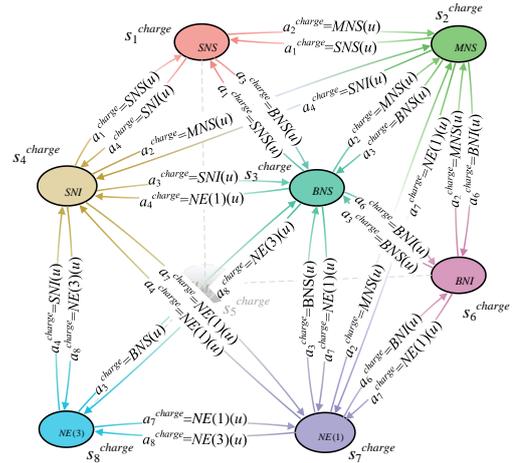

Fig.1. A fully connected digraph state-action network of the independent learning model

(3) **CM based Reward function**: Reward function $r(s, a)$ is essentially used to reinforce the effective action for each state. In the minimization-based problems. A so-called effective action can reduce the fitness of a feasible best solution or improve the coupling of a loosely coupled solution. Thus, the immediate reward of applying action $s$ in a certain state, denoted as $r(s, a)$ is determined by the contribution of an action for enhancing the coupling or improving the fitness value of the current solution.

$$r(s,a) = \begin{cases} 1.5 & \Delta f(\pi) < 0 \text{ and } \Delta CM(\pi) > 0 \\ 1 & \Delta f(\pi) < 0 \text{ and } \Delta CM(\pi) \leq 0 \\ 0.2 & \Delta f(\pi) \geq 0 \text{ and } \Delta CM(\pi) > 0 \\ 0 & \text{otherwise} \end{cases} \quad (22)$$

(4) **Adaptive selection mechanism of actions**: The $\varepsilon_t$-greedy selection strategy is used to choose a stochastic action with a probability $\varepsilon$ or an action possess the highest Q-value under a candidate state with a probability 1-$\varepsilon_t$, which is proposed in Eq. (24).

$$a_t = \begin{cases} 1. \text{Random select, if } rand[0,1] < \varepsilon_t \\ 2. \arg\max_{a \in A} Q(s_t, a), \text{otherwise} \end{cases} \quad (23)$$

$$SelectAnAction(Current\text{-}state, \varepsilon_t) \quad (24)$$

The actions selection mechanism (see Eq. (24)) is intended to regulates the equilibrium between exploitation and exploration in the learning process. In the early iterations of the learning process, the current solution is often far from the optimal solution. A larger value of $\varepsilon_t$ is conducive for exploring a large-scale solution space to find a promising region. Contrarily, with the extension of the running time of the algorithm, the obtained solution gradually approaches the optimal solution, and a small $\varepsilon$ value is in favor of an intensive search. Therefore, we design an adaptive adjustment dynamically controlled mechanism for $\varepsilon_t$, as shown in Eq. (25):

$$\varepsilon_t = (\varepsilon_0 - \varepsilon_{max}) \times \frac{T_{total} - T_{current}}{T_{total}} - \varepsilon_{max} \quad (25)$$

Where $\varepsilon_0$ is a large initial value for $\varepsilon_t$, $\varepsilon_{max}$ is a small final value for $\varepsilon_t$, $T_{total}$ stands for the total execution time, $T_{current}$ donates the execution time.

(5) **Updating Q-table**: The Q-table is responsible for saving the reinforcement signals of the state-action pairs during the learning process, each value of $Q(s_i,a_j)$ in the Q-table is used to reflect the search ability of the $j$th action at $s_i$. Therefore, each element of this table is set as 0 to ensure the departure point of the learning process is unbiased. The initialized Q-table with dimensions $M \times M$ is shown in Eq.(26), where $M$ is the number of states. Subsequently, $Q(s_i,a_j)$ is updated by adding its immediate reward to the original one by using the linear weighting method. The one-step updating rule of the Q-value is shown as Eq.(27).

$$Q\_table_{initial} = \begin{array}{c} a_1 \\ a_2 \\ \ldots \\ a_M \end{array} \begin{bmatrix} s_1 & s_2 & \ldots & s_M \\ 0 & 0 & \ldots & 0 \\ 0 & 0 & \ldots & 0 \\ \ldots & \ldots & \ldots & \ldots \\ 0 & 0 & 0 & 0 \end{bmatrix}_{M \times M} \quad (26)$$

$$Q(s_i, a_j) = (1-\alpha)Q(s_i, a_j) + \alpha r(s_i, a_j) \quad (27)$$

### C. QLSF-based L2I tier

In the L2I tier, three cooperative Q-learning-based local search frameworks (QLSFs)—namely, Charge_QLSF, Cast_QLSF, and SQLSF—are constructed. The first two, Charge_QLSF and Cast_QLSF, are designed as independent local search frameworks. They aim to improve the solution to one subproblem while disregarding the state of the other. In contrast, SQLSF is developed to learn the joint state of both the charge and cast subproblems, enabling the selection of an optimal joint action that simultaneously enhances both subsequences of the original problem. To further accelerate the neighborhood search process, a validity evaluation method is proposed for Charge_QLSF to eliminate ineffective improvement operators, while a speed-up evaluation method is introduced for Cast_QLSF.

*1) Charge_QLSF*

The Charge_QLSF begins with a feasible solution $\pi = (u, v)$, where u and v represent the charge and cast subsequences, respectively. It then seeks an optimal neighborhood operator to improve the charge subsequence u, assuming the cast agent is out of service (i.e., v remains fixed). As detailed in **Algorithm 1**, the *Charge_QLSF* initializes all elements of the Qcharge-table to zero (Lines 1 and 3) to ensure an unbiased starting point. It then iteratively applies neighborhood operators selected via the $\varepsilon_t$-greedy strategy (Line 7) to guide the search towards an optimal or near-optimal solution (Lines 8 and 18). During this process, the Qcharge-table is updated based on the rewards received, and a new state is recorded to initiate the next round of iterative improvement.

Furthermore, historical observations indicate that calculating the target values for neighboring solutions is time-consuming and can hinder the efficiency of the downward search. To address this issue, a validity evaluation method—grounded in the general properties of the permutation-based model for the SCCSP, as discussed in Section II—is introduced. This method filters out invalid improvement operations, thereby reducing the computational complexity of the HierC_Q algorithm.

Denote $\pi_{j_1}$ and $\pi_{j_2}$ the selected two charges from the charge set $\{\pi_1 \ldots \pi_{j_1} \ldots \pi_{j_2} \ldots \pi_{\alpha_i}\}$ of $\pi_i^{cast}$ to excuse the improvement operation $a_i^{charge}$. $u'$ the charge subsequence after performing the improvement operation. $Pri(v, j_1, j_2) = j_1 \succ j_2$ the processing priority of $\pi_{j_1}$ is higher than that of $\pi_{j_2}$ in the cast subsequence $v$. $Pri(u', j_1, j_2)$ the processing priority of charge $\pi_{j_1}$ and $\pi_{j_2}$ in the charge subsequence after performing the improvement operation. $L(\pi_i^{cast(k)})$ and $L'(\pi_i^{cast(k)})$ the start processing time of cast $\pi_i^{cast}$ on $M_k$ before and after performing $a_i^{charge}$ separately. $R_k$ and $R'_k$ the completion time of cast $\pi_i^{cast}$ on $M_k$ before and after executing $a_i^{charge}$ separately. The validity evaluation theorem of improvement operator $a_i^{charge}$ is as follows:

**Theorem 1**: Suppose that two charges $\pi_{j_1}$ and $\pi_{j_2}$ are selected from the same cast $\pi_i^{cast}$, and they need to be executed $a_i^{charge}$. If $Pri(u', j_1, j_2) \neq Pri(v, j_1, j_2)$, $a_i^{charge}$ is invalid.

The proof of Theorem 1 is provided in **Part 4** of Supplementary Material for completeness.

| Algorithm 1 Charge_QLSF |
|---|
| **Input:** Candidate Solution $\pi=(u, v)$ and the episode period $Ep_{charge}$ |
| **Output:** $\pi_{best}^{charge}$ |
| 1: **If** the first time of using the Charge_QLSF **Then** |
| 2:   Initialize the $Q_{charge}$-table of all state-action pairs to 0 |
| 3: **End If** |
| 4: $\pi'(u', v')=\pi(u, v)$, $t=1$, $count =0$, randomly determine the initial state $s_i^{charge}$ from $\Sigma_{charge}$ |
| 5: Calculate $\varepsilon_{Ep}$ according to Eq. (26) |
| 6: **While**($t<Ep_{charge}$) **do** |
| 7:   $a_i^{charge}=SelectAnAction(s_i^{charge}, \varepsilon_{Ep})$ // Select an action from $A_{charge}$ |
| 8:   **Repeat** |
| 9:     $u''= a_i^{charge}(u')$. //Perform $a_i^{charge}$ on the charge subsequence $u'$ of $\pi'$ at $s_i^{charge}$ |

10:   If $a_i^{charge}$ is invalid **Then** *count= count*+1
11:   **Else**
12:       Combine $u''$ and $v'$ into $\pi''(u'', v')$ and calculate the fitness of $\pi''$
13:       Calculate P$_{charge}$=$r(s_i^{charge}, a_i^{charge})$ according to Eq.(22)
14:       **If** P$_{charge}$=$r(s_i^{charge}, a_i^{charge})$>0 **Then**
15:           $\pi'$=$\pi''$, *count* =0, $Q_{charge}(s_i^{charge}, a_i^{charge})$=$\alpha Q_{charge}(s_i^{charge}, a_i^{charge})$+(1-$\alpha$) $r(s_i^{charge}, a_i^{charge})$
16:       **Else** *count = count* +1
17:       **End If**
18:   **End If**
19:   **Until** *count>N*
20:   **If** $f(\pi'') < f(\pi')$ **Then** *t*=0
21:   **Else** *t= t*+1
22:   **End If**
23:   Perceive new state $s_i^{charge}$ from the state set $\Sigma_{charge}$
24: **End While**
25: **Return** $\pi_{best}^{charge}$

*2) Cast_QLSF*

Analogous to the Charge_QLSF, the Cast_QLSF focuses on identifying an optimal neighborhood operator to enhance the cast subsequence $v$, while keeping the charge subsequence $u$ fixed. The detailed procedure of Cast_QLSF is outlined in **Algorithm 2**. To improve computational efficiency, a speed-up evaluation method is also developed based on the general properties of the permutation-based model. This method aims to accelerate the neighborhood search process by reducing unnecessary evaluations and focusing on promising candidate solutions.

Denote $R(\pi_j^{cast(k)})$ is the release time of $\pi_j^{cast(k)}$ when arriving at the continuous casting stage. $L(\pi_j^{cast(k)})$ the starting time of $\pi_j^{cast(k)}$. $C(\pi_j^{cast(k)})$ the completion time of $\pi_j^{cast(k)}$. $R_k$ the final release time on $M_k$. $C_{max} = \max\{R_a | a=1,...,m_S\}$ the maximum completion time. $\pi_{j_{key}}^{cast(k)}$ the critical cast on machine $M_k$, where

$$j_{key} = \max_{j=1,2...T_S(k)}\left\{j\left|L\left(\pi_j^{cast(k)}\right)-R\left(\pi_j^{cast(k)}\right)=0\right.\right\}.$$ $L'(\pi_j^{cast(k)})$ the starting time of $\pi_j^{cast(k)}$ after performing $a_i^{cast}$. $\pi_{j'_{key}}^{cast(k)}$ the key cast on $M_k$ after executing $a_i^{cast}$. $R'_k$ the final release time of $M_k$ after performing $a_i^{cast}$. $C'_{max} = \max\{\max\{R_a | a=1,...,m_S, a \neq k\}, R'_k\}$ the makespan (i.e., the maximum completion time) of the new neighbor or solution. Since the cast subproblem is actually a parallel machine scheduling problem, we denote the critical machine at the continuous casting stage as the machine with the largest completion time (i.e., the machine with the biggest final release time).

*Theorem 2*: Suppose that two casts are processed on two different machines $M_{k_1}$ and $M_{k_2}$, and they need to be performed $a_i^{cast}$. If there is at least one critical machine between $M_{k_1}$ and $M_{k_2}$, $L(\pi_{j'_{key}}^{cast(k_1)})+\sum_{j=j'_{key}}^{T_S(k_1)}p(\pi_j^{cast(k_1)})-L(\pi_{j_{key}}^{cast(k_1)})-\sum_{j=j_{key}}^{T_S(k_1)}p(\pi_j^{cast(k_1)}) < 0$ and $L(\pi_{j'_{key}}^{cast(k_2)})+\sum_{j=j'_{key}}^{T_S(k_2)}p(\pi_j^{cast(k_2)})-L(\pi_{j_{key}}^{cast(k_2)})-\sum_{j=j_{key}}^{T_S(k_2)}p(\pi_j^{cast(k_2)}) < 0$, then is has $C'_{max} < C_{max}$.

If $k_1 = k_2$ (i.e., $\pi_{j_1}^{cast(k_1)}$ and $\pi_{j_2}^{cast(k_2)}$ are processed on the same machine $M_k$), $M_k$ is critical machine, and $L(\pi_{j'_{key}}^{cast(k)})+\sum_{j=j'_{key}}^{T_S(k)}p(\pi_j^{cast(k)}) < L(\pi_{j_{key}}^{cast(k)})-\sum_{j=j_{key}}^{T_S(k)}p(\pi_j^{cast(k)})$, then it has $C'_{max} < C_{max}$.

Please refer to Supplementary Material **Part 4** for the detailed proof of Theorem 2.

**Algorithm 2** Cast_QLSF
**Input:** Candidate Solution $\pi=(u, v)$ and the episode period $Ep_{cast}$
**Output:** $\pi_{best(Ep)}^{cast}$
1: **If** the first time of using the *Cast_QLSF* **Then**
2:    Initialize the $Q_{cast}$-table of all state-action pairs to 0
3: **End If**
4: $\pi'(u', v')= \pi(u, v)$, *t*=1, *count* =0, randomly determine the initial state $s_i^{cast}$ from the state set $\Sigma_{cast}$
5: Calculate $\varepsilon_{Ep}$ according to Eq.(25)
6: **While**(*t<Ep$_{cast}$*) **do**
7:   $a_i^{cast}$=*SelectAnAction*($s_i^{cast}$, $\varepsilon_{Ep}$) // Select an action from A$_{cast}$
8:   **Repeat**
9:      $v''= a_i^{cast}(v')$. //Perform action $a_i^{cast}$ on the charge subsequence $v'$ of $\pi'$ at $s_i^{cast}$
10:     **If** $a_i^{caste}$ is invalid **Then** *count= count*+1
11:     **Else**
12:         Combine $u''$ and $v'$ into a new solution $\pi''(u'', v')$ and evaluate the fitness of $\pi''(u'', v')$ according to the proposed Theorem 2.
13:         Calculate P$_{cast}$=$r(s_i^{cast}, a_i^{cast})$ according to Eq.(22)
14:         **If** P$_{cast}$=$r(s_i^{cast}, a_i^{cast})$>0 **Then**
15:             $\pi'$=$\pi''$, *count* =0, $Q_{cast}(s_i^{cast}, a_i^{cast})$=$\alpha Q_{cast}(s_i^{cast}, a_i^{cast})$+(1-$\alpha$) $r(s_i^{cast}, a_i^{cast})$
16:         **Else** *count = count* +1
17:         **End If**
18:     **End If**
19:   **Until** *count>N*
20:   **If** $f(\pi'') < f(\pi')$ **Then** *t*=0
21:   **Else** *t= t*+1
22:   **End If**
23:   Perceive new state $s_i^{cast}$ from the state set $\Sigma_{cast}$
24: **End While**
25: **Return** $\pi_{best}^{cast}$

*3) SQLSF*

Distinct from the aforementioned independent QLSFs, the SQLSF employs aggregation techniques to construct joint action and state spaces. During the iterative improvement process, an improvement operator pair $a_i^{joint}$=($a_i^{cast}$, $a_i^{charge}$) is selected based on a corresponding joint state pair $s_i^{joint}$=($s_i^{cast}$, $s_i^{charge}$), thereby enabling the simultaneous optimization of both subsequences. The detailed procedure of SQLSF is described in **Algorithm 3**, where $v''= a_i^{cast}(v')$ and $u''= a_i^{charge}(u')$ denote the concurrent execution of improvement operations on the cast and charge subsequences, respectively. The iteration process is governed by the hyperparameter $Ep_{joint}$ which controls the number of iterations in SQLSF. As shown in Algorithm 3, HierC_Q initiates the search from promising regions of the solution space (i.e., candidate solution $\pi=(u, v)$), and iteratively applies the improvement operators (see Lines 8 and 17) to guide the search towards an optimal or near-optimal solution.

**Algorithm 3** SQLSF
**Input:** Candidate Solution $\pi=(u, v)$ and the episode period $Ep_{joint}$
**Output:** $\pi_{best}^{joint}$
1: **If** the first time of using the SQLSF **Then**
2:    Initialize the $Q_{joint}$-table of all state-action pairs to 0
3: **End If**
4: $\pi'(u', v')= \pi(u, v)$, *t*=1, *count* =0, randomly determine the initial state $s_{joint}$=($s_i^{cast}$, $s_i^{charge}$) from the joint state set $\Sigma_{joint}$
5: Calculate $\varepsilon_{Ep}$ according to Eq.(25)
6: **While**(*t<Ep$_{cast}$*) **do**
7:   $a_i^{joint}$=($a_i^{cast}$, $a_i^{charge}$) =*SelectAnAction*($s_i^{joint}$, $\varepsilon_{Ep}$) // Select an action the

from the joint action set $A_{joint}$
8: **Repeat**
9:    $v''= a_i^{cast}(v')$ //Perform action $a_t^{cast}$ on the cast subsequence $v'$ of $\pi'$ at $s_t^{cast}$
10:    $u''= a_i^{charge}(u')$ //Perform action $a_t^{charge}$ on the charge subsequence $u'$ of $\pi'$ at $s_t^{charge}$
11:    Combine $u''$ and $v''$ into a new solution $\pi''(u'', v'')$ and evaluate the fitness of $\pi''(u'', v'')$
12:    Calculate $P_{joint}=r(s_i^{joint}, a_i^{joint})$ according to Eq.(22)
13:    **If** $P_{joint}= r(s_i^{joint}, a_i^{joint})>0$ **Then**
14:      $\pi'=\pi''$, $count =0$, $Q_{joint}(s_i^{joint}, a_i^{joint})=\alpha Q_{joint}(s_i^{joint}, a_i^{joint})+(1-\alpha)r(s_i^{joint}, a_i^{joint})$
15:    **Else** $count = count +1$
16:    **End If**
17: **Until** $count>Z$
18: **If** $f(\pi'')< f(\pi')$ **Then** $t=0$
19: **Else** $t= t+1$
20: **End If**
21: Perceive new state $s_i^{joint}$ from the state set $\Sigma_{joint}$
22: **End While**
23: **Return** $\pi_{best}^{joint}$

### D. D2R tier

The D2R (Disturb-to-Renovate) tier is responsible for introducing solution perturbations to maintain search diversity and prevent premature convergence. Two key challenges are addressed: when to trigger perturbation and how to execute it. To this end, we propose a construction-and-perturbation-based renewal strategy.

A control parameter, $\gamma$, is introduced to determine when the D2R tier is activated. If the best historical objective value fails to improve over $\gamma$ consecutive iterations, the algorithm is assumed to be trapped in a local optimum, thus triggering the renewal process.

Inspired by the observation that elite solutions often contain valuable structures [10], the renewal strategy combines constructive heuristics with two custom perturbation operators to generate diverse candidates::

1) *Insert_f(v)*: Randomly selects a position in the cast subsequence $v$ and moves the subsequence following this position to the front.

2) *Inter_r(v)*: Randomly selects two positions in $v$ and reverses the subsequence between them, ensuring their distance $d(a,b)>Z/3$ for adequate diversification.

These perturbations enhance search diversity and facilitate effective escape from local optima. The renewal strategy (see **Algorithm 5**) randomly applies one of the two operators to the cast subsequence, then reconstructs the corresponding charge subsequence using a heuristic (**Algorithm 4**), resulting in a structurally distinct solution.

Empirical results confirm that restarting the L2I phase from such diversified solutions effectively re-invigorates the search, even if the new solution is initially inferior. This mechanism demonstrates strong capability for escaping local optima and maintaining search vitality.

**Algorithm 4** *u_constructing procedure*
**Input:** Cast subsequence $v$
**Output:** Charge subsequence $u$
1: Set $j=1$
2: **While**($j \leq Z$) **do**
3:    Schedule the casts to the first available machine according to $v$ in the continuous casting stage
4:    Compute the starting processing time for each charge of each cast
5: **End While**
6: Sort all of the charges in an increasing order of the start processing time, and obtain a charge subsequence $u$
7: **Return** The generated charge subsequence $u$

**Algorithm 5** The construction-and-perturbation based renewal strategy
**Input:** Current solution $\pi=(u, v)$
**Output:** Reconstruction solution $\pi'=(u', v')$
1: **If** $r=random[0,1] >0.5$
2:    $v'=Inter\_r(v)$
3: **Else**
4:    $v'=Insert\_f(v)$
5: **End If**
6: Generate a new charge subsequence $u'$ according to $u\_constructing$ procedure($v'$)
7: Combine $u'$ and $v'$ into a new individual $\pi'$
8: **Return** $\pi'=(u', v')$

### E. The architecture of HierC_Q

**Algorithm 6** summarizes the overall workflow of HierC_Q, which integrates three cooperative Q-learning-based local search frameworks (QLSFs) and a construction-and-perturbation-based renewal strategy. The process begins by initializing parameters and Q-tables, generating the initial cast subsequence $v_0$ via the LPT rule, and constructing the initial charge subsequence $u_0$ (Algorithm 4) to form the initial solution $\pi_0=(u_0, v_0)$. The best solution $\pi_{best}$ is set accordingly.

In each iteration, Charge_QLSF, Cast_QLSF, and SQLSF (**Algorithms 1–3**) are invoked in sequence to refine the current solution $\pi_{gen}$. Any improvement is accepted as the new $\pi_{gen}$, and if it also surpasses $\pi_{best}$, the best solution is updated and the improvement counter is reset; otherwise, the counter increments. When no improvement is observed over $\gamma$ consecutive iterations, the D2R strategy (**Algorithm 5**) is triggered to generate a new starting point. The procedure repeats until the termination criterion is met, after which the best solution found is returned.

A detailed computational complexity analysis, provided in the supplementary material (**Part 5**), demonstrates that the dominant term is quadratic, indicating that HierC_Q achieves a good balance between solution quality and computational efficiency.

**Algorithm 6** HierC_Q procedure
**Input:** The iteration number $\gamma$
**Output:** Best solution $\pi_{best}$
1: $gen=0$
2: *LPT* rule is used to generate the initial cast subsequence $v_0$
3: $u_0 \leftarrow u\_constructing\ procedure(v_0)$, $\pi_0=(u_0, v_0)$// Generate $u_0$ based on **Algorithm 4**, and obtain the initial solution $\pi_0$
4: Calculate the fitness value of $\pi_0$, let $Count=0$, $\pi_{best}= \pi_0$
5: **Repeat**
6:   **While** ($Count<\gamma$) **do**
7:     $\pi_{best}^{charge} \leftarrow Charge\_QLSF(\pi_{gen})$ //Call **Algorithm 1** to output a new solution and calculate its fitness value
8:     **If** $f(\pi_{best}^{charge}) <f(\pi_{gen})$ **Then** $\pi_{gen} \leftarrow \pi_{best}^{charge}$
10:     **End If**
11:     $\pi_{best}^{cast} \leftarrow Cast\_QLSF(\pi_{gen})$ //Call **Algorithm 2** to output a new solution and calculate its fitness value
12:     **If** $f(\pi_{best}^{cast}) < f(\pi_{gen})$ **Then** $\pi_{gen} \leftarrow \pi_{best}^{cast}$
14:     **End If**
15:     $\pi_{best}^{joint} \leftarrow SLF(\pi_{gen})$//Call **Algorithm 3** to output a new solution and calculate its fitness value
16:     **If** $f(\pi_{best}^{joint}) < f(\pi_{gen})$ **Then** $\pi_{gen} \leftarrow \pi_{best}^{joint}$
18:     **End If**
19:     **If** $f(\pi_{gen}) < f(\pi_{best})$ **Then**
20:       $\pi_{best}= \pi_{gen}$, $Count=0$
21:     **Else**
22:       $Count= Count+1$
23:     **End If**
24:     $gen=gen+1$
25:   **End While**
26:   $\pi_{gen} \leftarrow D2R(\pi_{gen})$// Call **Algorithm 5** to generate a new solution
27: **Until** (Termination condition is met)
28: **Return** The best solution $\pi_{best}$

## V. COMPUTATIONAL COMPARISONS AND STATISTICAL ANALYSES

### A. Experimental Setup

Since there are no established benchmarks for the considered problem, a set of randomly generated instances inspired by real-world steelmaking complex scenarios is used to evaluate the performance of HierC_Q. Specifically, twenty combinations of $\{S \times Z\}$, where $S \in \{3,4,5,6\}$ and $Z \in \{10,15,20,25,30\}$ are considered. The processing time $p_{j \times k}$ and the setup time $q_j$ are randomly generated from a uniform distribution in the range [36,50] and a uniform distribution in the range [80,100], respectively. Other production data for the instances are generated by a uniform distribution in the following range: $m_i \in [3,5]$, $\alpha_j \in [8,12]$, $t_i \in [10,15]$. All of the testing instances and the supplementary material can be downloaded from the website (i.e., https://github.com/ly726564418/TCYBE.git).

Performance is measured by the average relative percentage deviation (ARPD), defined as:

$$ARPD(\pi) = \frac{1}{R}\sum_{i=1}^{R}(\frac{f_i(\pi) - f_{best}(\pi)}{f_{best}(\pi)}) \times 100\% \quad (28)$$

where $f_i(\pi)$ is the result from the $i$-th run, and $f_{best}(\pi)$ is the best among all compared algorithms for the same instance [30]. Lower ARPD values indicate better performance.

All algorithms are implemented in Delphi 2010 and executed on a standard PC (Intel i7-8550U, 1.8 GHz, 32 GB RAM, Windows 10). Each is run 30 times independently per instance, with a time limit $T_{total} > Z \times S \times \lambda$ ms as the stopping criterion, where $\lambda$ is the runtime factor. The best, second-best, and third-best results in each row of Tables II–III are marked accordingly.

Due to space constraints, only the main comparative results are presented here; full details, including parameter calibration, ablation studies (CM-based reward, cooperative QLSFs, PDNOs, perturbation-renewal), and complete results, are available in the supplementary material.

### B. Comparison of HierC_Q with eleven local search-based frameworks

This section presents the comparative results of HierC_Q against several local search frameworks (LSFs) and their variants, in order to verify the effectiveness of the proposed framework and the developed problem-dependent neighborhood operators (PDNOs). Five representative LSFs are considered: variable neighborhood search (VNS) [19], iterated local search (ILS) [20], great deluge algorithm (GD) [31], adaptive large neighborhood search (ALNS) [21], and hyper-heuristic genetic algorithm (HHGA) [22]. All of these frameworks are equipped with the PDNOs and perturbation operators introduced in this study to ensure a fair and thorough comparison. In the scheduling literature, Swap, Insert, and Exchange are classical neighborhood operators for permutation-based representations. To further validate the effectiveness of the proposed PDNOs, we implement variant versions of each algorithm—namely, VNS$_{var}$, ILS$_{var}$, GD$_{var}$, ALNS$_{var}$, HHGA$_{var}$, and HierC_Q$_{var}$—that employ these traditional operators. The pseudocode implementations of these frameworks for solving the SCCSP problem are provided in **Part 7** of the Supplementary Material, accessible via the previously mentioned website.

Statistical results under three runtime factors, $\lambda$=200, 300, and 400, are reported in the **Part 6** of the Supplementary Material. Mean plots with 95% Tukey's HSD confidence intervals for HierC_Q and the eleven compared algorithms are also provided therein. Table II lists the results for $\lambda$=200. From both the results in the supplementary material and Table II, it is evident that HierC_Q consistently achieves lower ARPD values than all other algorithms under the same $\lambda$, across all instances. Furthermore, HierC_Q under $\lambda$=200 still outperforms most of the other algorithms even under $\lambda$ = 400, demonstrating its robustness.

Fig.7 and 8 in the Supplementary Material show group mean plots with 95% Tukey's HSD confidence intervals. Fig.6 illustrates that HierC_Q is statistically superior to all other algorithms under the same $\lambda$, regardless of whether traditional neighborhood operators (*Swap*, *Insert*, and *Exchange*) or the proposed PDNOs are used. This confirms the superior performance of HierC_Q over existing LSFs. Additionally, as shown in the subplots of Fig.7, LSFs equipped with PDNOs consistently outperform their traditional-operator counterparts under the same $\lambda$, indicating that PDNOs offer a powerful search mechanism for optimizing the SCCSP.

### C. Comparison HierC_Q with the state-of-the-art methods

To evaluate the effectiveness of HierC_Q, this section compares its performance with several state-of-the-art algorithms reported in recent literature. These algorithms are categorized into three groups.

The first group includes an industrial heuristic (IDH), which is widely adopted in modern iron and steel enterprises. IDH addresses the cast subproblem and the charge subproblem separately, solving the cast subproblem first. In the cast subproblem, all casts are sorted in descending order of processing time to form a cast subsequence. Each cast is then selected sequentially from the end to the beginning of this subsequence and assigned to the earliest available casting machine. This procedure establishes a partial scheduling plan by determining both the start processing time and the assigned casting machine for each cast. Simultaneously, the start processing times of the corresponding charges in the continuous casting stage are recorded. For the charge subproblem, all charges are sorted in ascending order of their recorded start times in the continuous casting stage to create a charge subsequence. Charges are then selected from the end to the beginning of this subsequence and sequentially assigned to available refining or steelmaking machines, moving backward from the final refining stage to the steelmaking stage. At each stage, charges are allocated to the first available machine and scheduled to start as late as possible. This process yields the complete scheduling plan.

The second group comprises four algorithms—PIDE [32], HFOA [33], IABC [8], and CCABC [10]—which are specifically designed to solve various types of steelmaking–continuous casting scheduling problems (SCCSPs). The third group consists of four algorithms—ISA [34], HGA [35], IPSO [36], and EIGA [37]—which are developed for hybrid flowshop scheduling problems.

All algorithms are thoroughly reimplemented in accordance with their original literature and are appropriately adapted to address the problem at hand. The corresponding pseudocodes, detailed descriptions, and parameter configurations for these algorithms are provided in **Part 8** of the Supplementary Material. The comprehensive comparative results under $\lambda$=200, 400, and 600 are presented in Tables X–XII of the **Part 6** in the Supplementary Material. Additionally, mean plots with 95% Tukey's HSD confidence intervals and the corresponding box plots for all test results under each instance are included in the same section. In the main text, the results under $\lambda=200$ and the performance histograms are reported in Table III and Fig.2, respectively.

From both the results in the Supplementary Material and Table III and Fig.2, it is evident that HierC_Q consistently achieves lower ARPD and standard deviation (SD) values than the compared algorithms across all instances. Moreover, HierC_Q under $\lambda=200$ outperforms nearly all competing algorithms even under $\lambda=400$, indicating its robustness and efficiency. These numerical results validate the strong problem-solving capability of the proposed HierC_Q for the SCCSP.

Furthermore, HierC_Q attains the lowest overall average values of ARPD and SD across 20 instances of varying scales, demonstrating superior average performance compared to the other nine algorithms. As reported in Subsections V-B and V-C, HierC_Q achieves the best results on the majority of instances while requiring only one-third of the computational time needed by the other methods. This performance advantage is primarily attributed to the CM-based reward function, which effectively guides the search process toward promising regions, and the cooperative QLSFs that leverage carefully designed neighborhood operators and problem-specific features to enhance exploitation efficiency. Therefore, HierC_Q can be regarded as an effective and efficient solution approach for the SCCSP.

Furthermore, **Part 9** of the Supplementary Material presents the best-performing results and associated Gantt charts generated by HierC_Q for 20 SCCSP instances of different scales.

TABLE II COMPARISON RESULTS OF HIERC_Q WITH ELEVEN HF_NSS UNDER $\Lambda=200$ AND 400.

| $S\times Z$ | $\lambda$ | VNS$_{var}$ | VNS | ILS$_{var}$ | ILS | GD$_{var}$ | GD | ALNS$_{var}$ | ALNS | HHGA$_{var}$ | HHGA | HierC_Q$_{var}$ | HierC_Q |
|---|---|---|---|---|---|---|---|---|---|---|---|---|---|
| 3×10 | | 0.755 | 0.758 | 0.727 | 0.799 | 0.721 | 0.753 | 0.818 | 0.81 | 1.678 | **0.588** | _0.641_ | **0.522** |
| 3×15 | | _0.131_ | **0.124** | 0.135 | 0.152 | **0.125** | 0.154 | 0.173 | 0.168 | 1.327 | 0.302 | 0.304 | 0.271 |
| 3×20 | | 0.834 | 0.820 | 0.862 | 0.699 | 0.677 | 0.606 | 0.662 | 0.773 | _0.546_ | 0.584 | **0.493** | 0.358 |
| 3×25 | | 1.166 | 1.044 | 1.182 | 0.993 | 0.874 | 0.950 | 1.288 | 0.919 | **0.457** | 0.767 | _0.536_ | 0.409 |
| 3×30 | | 1.596 | 1.481 | 1.614 | 1.610 | 1.620 | 1.632 | 1.615 | 1.632 | 0.187 | _0.926_ | **0.765** | 0.496 |
| 4×10 | | 1.338 | 1.427 | 1.230 | 1.303 | 1.005 | 1.460 | 1.118 | 1.424 | _0.690_ | 0.762 | **0.668** | 0.500 |
| 4×15 | | 0.447 | 0.449 | 0.471 | 0.536 | 0.474 | 0.491 | **0.089** | 0.084 | 0.546 | 0.409 | _0.343_ | 0.367 |
| 4×20 | | 0.616 | 0.611 | 0.325 | 0.348 | 0.332 | 0.391 | 0.274 | 0.298 | 0.679 | _0.209_ | **0.174** | 0.125 |
| 4×25 | | 0.939 | 1.100 | 1.100 | 1.020 | 0.849 | 0.900 | 1.015 | 0.977 | **0.580** | 0.853 | _0.686_ | 0.480 |
| 4×30 | | 0.715 | 0.806 | 0.824 | 0.627 | _0.516_ | 0.574 | 0.664 | 0.655 | **0.416** | 0.959 | 0.723 | **0.489** |
| 5×10 | 200 | 0.374 | 0.368 | _0.326_ | 0.383 | **0.273** | 0.372 | 0.464 | 0.397 | 0.688 | 0.354 | 0.360 | 0.246 |
| 5×15 | | 1.220 | 1.106 | 1.256 | 1.260 | 1.248 | 1.235 | 1.270 | 1.269 | 0.739 | _0.387_ | **0.390** | 0.206 |
| 5×20 | | 1.170 | 1.178 | 1.210 | 1.127 | 1.027 | 1.097 | 1.062 | _0.931_ | 1.187 | 0.964 | **0.754** | 0.423 |
| 5×25 | | 0.531 | 0.506 | 0.533 | 0.532 | 0.535 | 0.525 | 0.537 | 0.532 | 0.678 | **0.222** | _0.240_ | 0.154 |
| 5×30 | | 0.127 | **0.120** | **0.121** | 0.154 | _0.126_ | 0.155 | 0.167 | 0.156 | 0.519 | 0.333 | 0.318 | 0.304 |
| 6×10 | | 1.417 | 0.956 | 1.357 | 1.426 | 1.421 | 1.478 | 1.604 | 1.555 | 0.657 | _0.365_ | **0.309** | 0.220 |
| 6×15 | | 1.168 | 0.928 | 1.213 | 1.208 | 1.226 | 1.167 | 1.242 | 1.236 | 0.715 | _0.412_ | **0.345** | 0.230 |
| 6×20 | | 0.240 | _0.222_ | **0.217** | 0.256 | 0.214 | 0.247 | 0.338 | 0.334 | 0.840 | 0.333 | 0.308 | 0.248 |
| 6×25 | | 0.640 | 0.702 | 0.670 | _0.612_ | **0.581** | 0.624 | 0.671 | 0.71 | 0.702 | 0.642 | 0.630 | 0.481 |
| 6×30 | | 0.544 | 0.454 | 0.560 | 0.552 | 0.568 | 0.567 | 0.599 | 0.600 | 0.724 | _0.205_ | **0.197** | 0.127 |
| Average | | 0.798 | 0.758 | 0.797 | 0.780 | 0.721 | 0.769 | 0.783 | 0.773 | 0.728 | _0.529_ | **0.459** | 0.333 |

TABLE III COMPARISON RESULTS OF HIERC_Q WITH THE STATE-OF-THE-ART METHODS UNDER $\Lambda=200$.

| $S\times Z$ | IDH | ISA | | PIDE | | HGA | | HFOA | | IPSO | | EIGA | | IABC | | CCABC | | HierC_Q | |
|---|---|---|---|---|---|---|---|---|---|---|---|---|---|---|---|---|---|---|---|
| | ARPD | ARPD | SD | ARPD | SD | ARPD | SD | ARPD | SD | ARPD | SD | ARPD | SD | ARPD | SD | ARPD | SD | ARPD | SD |
| 3×10 | 2.319 | 1.479 | 0.478 | 1.513 | 0.463 | 2.361 | _0.346_ | 0.966 | 0.419 | **0.531** | **0.003** | 0.898 | 0.696 | 0.694 | 0.443 | _0.673_ | **0.336** | 0.522 | 0.522 |
| 3×15 | 1.372 | 0.417 | **0.143** | 0.417 | _0.192_ | 1.479 | **0.162** | 0.331 | 0.311 | 0.300 | 0.767 | **0.273** | 0.347 | 0.278 | 0.692 | _0.276_ | 0.766 | 0.271 | 0.306 |
| 3×20 | 3.427 | 1.644 | 5.243 | 1.482 | 5.623 | 0.698 | **0.154** | 1.100 | 6.256 | **0.366** | 0.448 | 0.651 | 3.436 | 0.358 | 3.339 | _0.561_ | 1.104 | 0.358 | 0.063 |
| 3×25 | 9.849 | 2.221 | 5.804 | 2.044 | 6.217 | _0.897_ | 2.876 | 1.600 | 4.963 | 1.799 | **1.510** | 1.731 | 3.285 | 1.701 | 1.893 | _0.674_ | _1.638_ | 0.409 | 0.376 |
| 3×30 | 3.352 | 1.582 | _1.079_ | 1.547 | 1.729 | 0.632 | 1.701 | 0.994 | 6.712 | 0.551 | **1.069** | 1.129 | 2.419 | 0.510 | **0.000** | 0.812 | 4.868 | 0.496 | 3.242 |
| 4×10 | 7.032 | 1.382 | 2.680 | 1.321 | 1.781 | 0.921 | **0.401** | 0.818 | 2.687 | 0.517 | 0.740 | _0.509_ | _0.630_ | **0.462** | 1.157 | 0.643 | 0.733 | **0.500** | 0.280 |
| 4×15 | 0.874 | 0.636 | **0.676** | 0.623 | _1.321_ | 0.674 | **0.000** | 0.494 | 1.355 | 0.380 | 1.689 | _0.378_ | 1.335 | 0.380 | 1.530 | **0.370** | 2.117 | **0.367** | 1.778 |
| 4×20 | 2.513 | 0.504 | 1.233 | 0.507 | 1.646 | 0.789 | _0.769_ | 0.372 | 1.712 | **0.127** | **0.627** | 0.224 | 1.289 | _0.149_ | 1.523 | 0.193 | 0.799 | **0.125** | _0.479_ |
| 4×25 | 10.508 | 2.450 | 5.022 | 2.395 | 3.789 | _0.882_ | 1.543 | 1.786 | 3.368 | 1.896 | **0.812** | 1.682 | 3.813 | 1.630 | 1.855 | **0.668** | _1.121_ | 0.480 | 0.437 |
| 4×30 | 2.218 | 1.384 | 1.385 | 1.410 | 0.663 | **0.499** | _0.534_ | 0.996 | 3.424 | 0.888 | **0.207** | 1.164 | 3.180 | _0.803_ | **0.464** | 0.733 | 2.850 | 0.489 | 1.620 |
| 5×10 | 4.327 | 1.323 | 2.612 | 1.302 | 2.136 | 1.162 | 1.126 | 0.678 | 2.806 | **0.272** | _0.275_ | **0.208** | 0.923 | 0.298 | 0.963 | 0.342 | 0.509 | 0.246 | **0.000** |
| 5×15 | 3.017 | 0.827 | 1.396 | 0.873 | 1.705 | 1.007 | 0.432 | 0.501 | 1.299 | _0.210_ | _0.858_ | **0.208** | 0.979 | 0.215 | **0.836** | 0.434 | 1.562 | 0.206 | 0.264 |
| 5×20 | 7.058 | 3.386 | 3.752 | 3.185 | 4.170 | _1.551_ | _2.087_ | 2.841 | 3.520 | 1.925 | 2.790 | 1.690 | 3.655 | 1.978 | 2.418 | **0.915** | _1.568_ | 0.423 | 0.236 |
| 5×25 | 2.365 | 0.477 | 1.435 | 0.480 | **0.845** | 0.796 | _1.020_ | 0.402 | 1.326 | **0.157** | 1.343 | 0.275 | 1.692 | _0.174_ | 1.354 | 0.247 | 1.303 | 0.154 | 0.350 |
| 5×30 | 0.729 | _0.294_ | 1.895 | **0.280** | 2.315 | 0.617 | **1.263** | **0.245** | 1.914 | 0.324 | 2.413 | 0.313 | 2.665 | 0.307 | 2.924 | 0.331 | _1.310_ | 0.304 | 0.523 |
| 6×10 | 3.279 | 0.786 | 1.631 | 0.840 | 2.227 | 0.864 | 0.627 | 0.366 | 2.547 | _0.247_ | 1.117 | 0.297 | 0.774 | **0.230** | **0.133** | 0.358 | **0.000** | 0.220 | _0.421_ |
| 6×15 | 3.011 | 0.782 | 0.989 | 0.765 | 1.848 | 1.045 | _0.795_ | 0.512 | 0.984 | 0.283 | 0.853 | **0.252** | **0.479** | _0.261_ | 1.426 | 0.413 | 1.482 | 0.230 | 0.170 |
| 6×20 | 1.477 | 0.924 | 1.372 | 0.923 | _1.264_ | 1.195 | 0.537 | 0.637 | 1.596 | **0.326** | 1.298 | 0.437 | 1.509 | 0.361 | 1.780 | _0.334_ | **0.911** | 0.248 | 1.266 |
| 6×25 | 8.692 | 2.507 | 6.451 | 2.529 | 6.430 | _0.823_ | _1.094_ | 1.956 | 3.707 | 1.378 | 2.639 | 1.030 | 3.074 | 1.175 | 1.766 | **0.632** | **0.928** | 0.481 | **0.000** |
| 6×30 | 1.773 | 0.540 | 1.986 | 0.533 | 3.362 | 0.921 | 2.777 | 0.344 | 2.586 | **0.133** | **0.000** | 0.239 | 2.691 | **0.127** | **0.859** | _0.187_ | 3.080 | 0.127 | _1.411_ |
| Average | 3.960 | 1.277 | 2.363 | 1.248 | 2.486 | 0.991 | **1.012** | 0.897 | 2.675 | 0.631 | _1.110_ | 0.683 | 1.944 | _0.605_ | 1.218 | **0.490** | 1.449 | 0.333 | 0.687 |

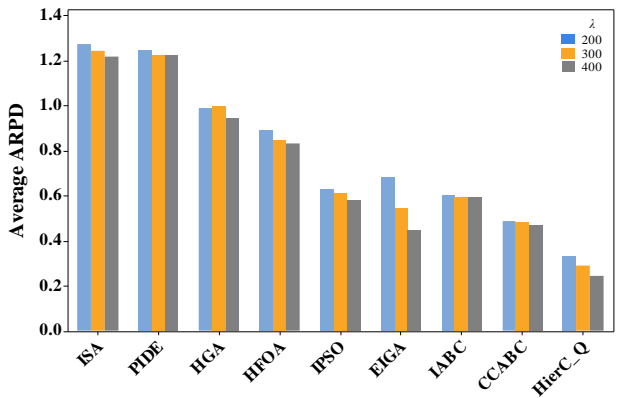 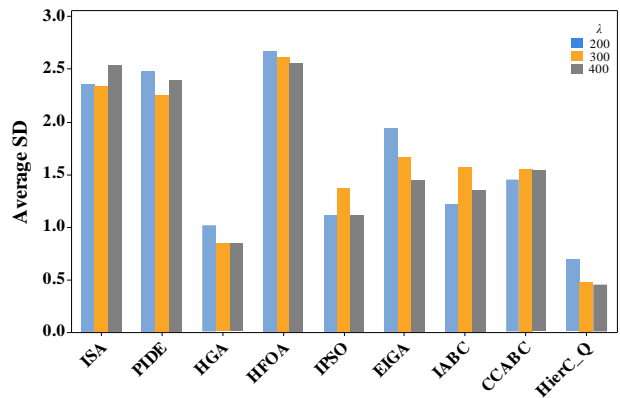

Fig2. Comparisons of HierC_Q's performance with the state-of-the-art algorithms.

## VI. CONCLUSION AND FUTURE RESEARCH

This paper presents a Q-learning-based hierarchical cooperative local search framework (HierC_Q) to address the steelmaking and continuous casting scheduling problem (SCCSP), a well-known decomposable scheduling challenge. HierC_Q integrates several key innovations: a coupling-measure (CM)-based reward function to guide exploration, three cooperative Q-learning-based local search frameworks (QLSFs) employing problem-dependent improvement operators for deep multi-neighborhood search, efficient validity and speed-up evaluation methods to reduce computational cost, and a construction-and-perturbation-based renewal strategy to maintain search diversity and prevent stagnation. Together, these components enable an effective balance between exploration and exploitation.

To our knowledge, this is the first work to apply a local search-based approach to the SCCSP. Moreover, the proposed framework—combining coupling measurement with cooperative local search—demonstrates broad applicability and can be readily extended to other decomposable scheduling problems, such as the assembly job shop scheduling problem (AJSSP) and the integrated scheduling problem (ISP). Given that the coupling measure captures essential features of discrete solution spaces, the insights and methodology developed here are potentially valuable for a wide range of combinatorial optimization problems.